\documentclass[lettersize,journal]{IEEEtran}
\usepackage{amsmath,amsfonts}
\usepackage{algorithmic}
\usepackage{array}
\usepackage[caption=false,font=normalsize,labelfont=sf,textfont=sf]{subfig}
\usepackage{textcomp}
\usepackage{stfloats}
\usepackage{url}
\usepackage{verbatim}
\usepackage{graphicx}

\usepackage{multirow} %自己加的
\usepackage{float}
\usepackage[colorlinks,
linkcolor=black,
anchorcolor=black,
citecolor=black]{hyperref}

\def\BibTeX{{\rm B\kern-.05em{\sc i\kern-.025em b}\kern-.08em
    T\kern-.1667em\lower.7ex\hbox{E}\kern-.125emX}}
\usepackage{balance}

\begin{document}
\title{Unsupervised and Interpretable Synthesizing for Electrical Time Series Based on Information Maximizing Generative Adversarial Nets}
\author{Zhenghao Zhou,~\IEEEmembership{Student Member,~IEEE,}
	Yiyan Li,~\IEEEmembership{Member,~IEEE,}
        Runlong Liu,~\IEEEmembership{Student Member,~IEEE,}
	Zheng Yan,~\IEEEmembership{Senior Member,~IEEE,}
	Mo-Yuen Chow,~\IEEEmembership{Fellow,~IEEE}

\thanks{
This work was supported by National Natural Science Foundation of China under Grant 52307121, and also supported by Shanghai Sailing Program under Grant 23YF1419000. (Corresponding author: Yiyan Li.)}
\thanks{
Zhenghao Zhou, Yiyan Li, Runlong Liu are with the College of Smart Energy, Shanghai Non-Carbon Energy Conversion and Utilization Institute, and Key Laboratory of Control of Power Transmission and Conversion, Ministry of Education, Shanghai Jiao Tong University, Shanghai, 200240, China. (e-mail: zhenghao.zhou@sjtu.edu.cn, yiyan.li@sjtu.edu.cn, runlong\_liu@sjtu.edu.cn).}
\thanks{
Zheng Yan is with the Key Laboratory of Control of Power Transmission and Conversion, Ministry of Education, and the Shanghai Non-Carbon Energy Conversion and Utilization Institute, Shanghai Jiao Tong University, Shanghai 200240, China. (e-mail: yanz@situ.edu.cn)}
\thanks{
Mo-Yuen Chow is with the University of Michigan - Shanghai Jiao Tong University Joint Institute, Shanghai Jiao Tong University, Shanghai, 200240, China. (email: moyuen.chow@sjtu.edu.cn)}
}

\markboth{Journal of \LaTeX\ Class Files,~Vol.~18, No.~9, September~2020}%
{how}

\maketitle

\begin{abstract}
Generating synthetic data has become a popular alternative solution to deal with the difficulties in accessing and sharing field measurement data in power systems. However, to make the generation results controllable, existing methods (e.g. Conditional Generative Adversarial Nets, cGAN) require labeled dataset to train the model, which is demanding in practice because many field measurement data lacks descriptive labels. In this paper, we introduce the Information Maximizing Generative Adversarial Nets (infoGAN) to achieve interpretable feature extraction and controllable synthetic data generation based on the unlabeled electrical time series dataset. Features with clear physical meanings can be automatically extracted by maximizing the mutual information between the input latent code and the classifier output of infoGAN. Then the extracted features are used to control the generation results similar to a vanilla cGAN framework. Case study is based on the time series datasets of power load and renewable energy output. Results demonstrate that infoGAN can extract both discrete and continuous features with clear physical meanings, as well as generating realistic synthetic time series that satisfy given features. 
\end{abstract}

\begin{IEEEkeywords}
Information maximizing generative adversarial nets, unsupervised learning, feature extraction, synthetic data.
\end{IEEEkeywords}

\section{Introduction}
\IEEEPARstart{A}{dequate} and high-quality dataset is the precondition for data-driven studies in power systems. However, due to concerns of energy security and user privacy, field measurement data of power systems cannot be easily obtained and shared by the academic community, which is adverse to the development of data-driven technologies. In this case, generating and utilizing synthetic data becomes an alternative solution. Synthetic data is defined as the imitated data derived from the field measurements, which has similar characteristics with the field measurements but does not correspond to any real-world individuals or objects. As a result, synthetic dataset can be accessed and shared by the academic community without additional concerns. 

In general, there are two main methods of generating synthetic data: simulation-based methods and data-driven methods. Simulation-based methods create synthetic data based on the simulation process of physical models, such as generating synthetic load profiles by load model simulation \cite{dickert2011time, gruber2012residential}. The advantages are that the results are explainable and controllable. However, the realisticness of the generated synthetic data relies heavily on the modeling accuracy and diversity, which is labor-intensive and inflexible. Data-driven methods generate synthetic data by learning the patterns from limited actual data, which is an end-to-end process and draws more attention due to its convenience and flexibility. 

Data-driven methods can be further divided into three categories: forecasting-based, clustering-based, and Generative Adversarial Nets (GAN)-based methods. Forecasting-based methods learn the patterns from historical data, and then generate synthetic data by projecting into the future. \cite{sarochar2019synthesizing} implements Long Short-Term Memory (LSTM) to learn patterns from historical energy consumption data, and then synthesized more data by utilizing the model’s forecasting ability. Forecasting-based methods are suitable for synthesizing time-series data, but the results lack diversity due to the supervised-learning nature. Clustering-based methods create synthetic data for unmeasured users by assigning them with field measurement data that comes from measurable users who fall into the same cluster, as illustrated in \cite{kim2011study}. However, clustering-based methods can hardly guarantee the realisticness because there are barely two users who have exactly the same characteristics. Starting from 2014, GAN-based methods have been widely studied in the field of synthetic data generation. 

GAN can learn the implicit distribution of a real-world dataset and generate unlimited amount of synthetic data that follow the same distribution, the results of which are realistic and diversified \cite{goodfellow2014generative}. \cite{zhang2018generative} utilizes GAN to learn the probability distribution of real datasets and generate samples based on this distribution, finding that the maximum mean discrepancy between real and synthetic samples tends to converge. \cite{el2020data} utilizes GAN and Kernel Density Estimators (KDE), to create synthetic datasets similar to real datasets that can be used in scenarios such as demand response, energy management, and non-intrusive load monitoring. \cite{zheng2022synthetic} proposes a method combining GAN and neural ordinary differential equations to generate high-fidelity synthetic phasor measurement unit data under varying load conditions, without needing to understand the underlying nonlinear dynamic equations. GAN-based methods can generate various types of data, not just load or sensor data. For instance, \cite{avkhimenia2021generation} explores the use of GAN to generate synthetic dynamic thermal line rating data and electricity pool prices. \cite{harell2021tracegan} introduces a GAN-based method called TraceGAN, which generates high-fidelity synthetic appliance power data. There are also many studies focusing on scenario generation. GAN-based methods can capture the uncertainty and spatiotemporal correlations of renewable energy plants without explicitly modeling the distributions and can generate diverse high-fidelity scenarios \cite{chen2018model,jiang2018scenario}. Considering privacy issues, \cite{li2021privacy} proposes a new framework called Fed-LSGAN, which combines federated learning and Least Squares Generative Adversarial Networks (LSGAN) to generate high-quality spatiotemporal scenarios of renewable energy while preserving privacy. Huang et al. introduce a model called Differentially Private Wasserstein Generative Adversarial Networks (DPWGAN), which combines the powerful generative power of generative adversarial networks with the strict privacy preservation of differential privacy \cite{huang2022dpwgan}.

To make the generation results controllable, conditional-GAN (cGAN) framework is further proposed to generate synthetic data upon certain conditions \cite{mirza2014conditional}. \cite{pinceti2021synthetic} encodes seasons and load types as labels to generate unique synthetic load profiles that match the labels. Yuan et al. proposed a method called C-StyleGAN2-SE for generating intraday renewable energy scenarios \cite{yuan2022conditional}. This method combines style-based generative adversarial networks with conditional generative adversarial networks, using meteorological variables as conditional inputs to generate high-fidelity renewable energy scenario data with complex spatiotemporal correlations. \cite{zhang2020typical} uses conditional improved generative adversarial networks to generate typical wind power scenarios for multiple wind farms, effectively capturing spatiotemporal correlations and improving the quality of the generated scenarios.

Note that the training process of cGAN requires labeled data (e.g. the user type, location, appliance name of a load profile). However, actual labels are even harder to acquire than measurement data due to user privacy and energy security concerns, while manual-labeling is costly and labor-intensive, which impair the applicability of cGAN in the energy sector. A commonly-used alternative solution is to first cluster the unlabeled data into different clusters, and then train vanilla GAN for each cluster respectively to enhance the generation results \cite{silva2023generating,gu2019gan,wang2020generating}. But the results still lack clear physical meanings and therefore are inexplainable. 

In order to address the shortcomings of the highly coupled and supervised learning, Information Maximizing Generative Adversarial Network (infoGAN) is proposed by Xi Chen et al. \cite{chen2016infogan}, which is an unsupervised learning method.

The biggest contribution of this paper is the development of time series-oriented infoGAN model, which has two advantages: unsupervised and interpretable feature extraction; integration of feature extraction and generation process to improve efficiency.

\section{Methodology}

 In this section, we first introduce the vanilla GAN and cGAN to serve as the basis and benchmark of the proposed infoGAN model. Then we introduce the infoGAN framework to achieve unsupervised and interpretable feature extraction and synthetic data generation. 

\begin{figure*}[!t]
	\centering
	\includegraphics[width=7in]{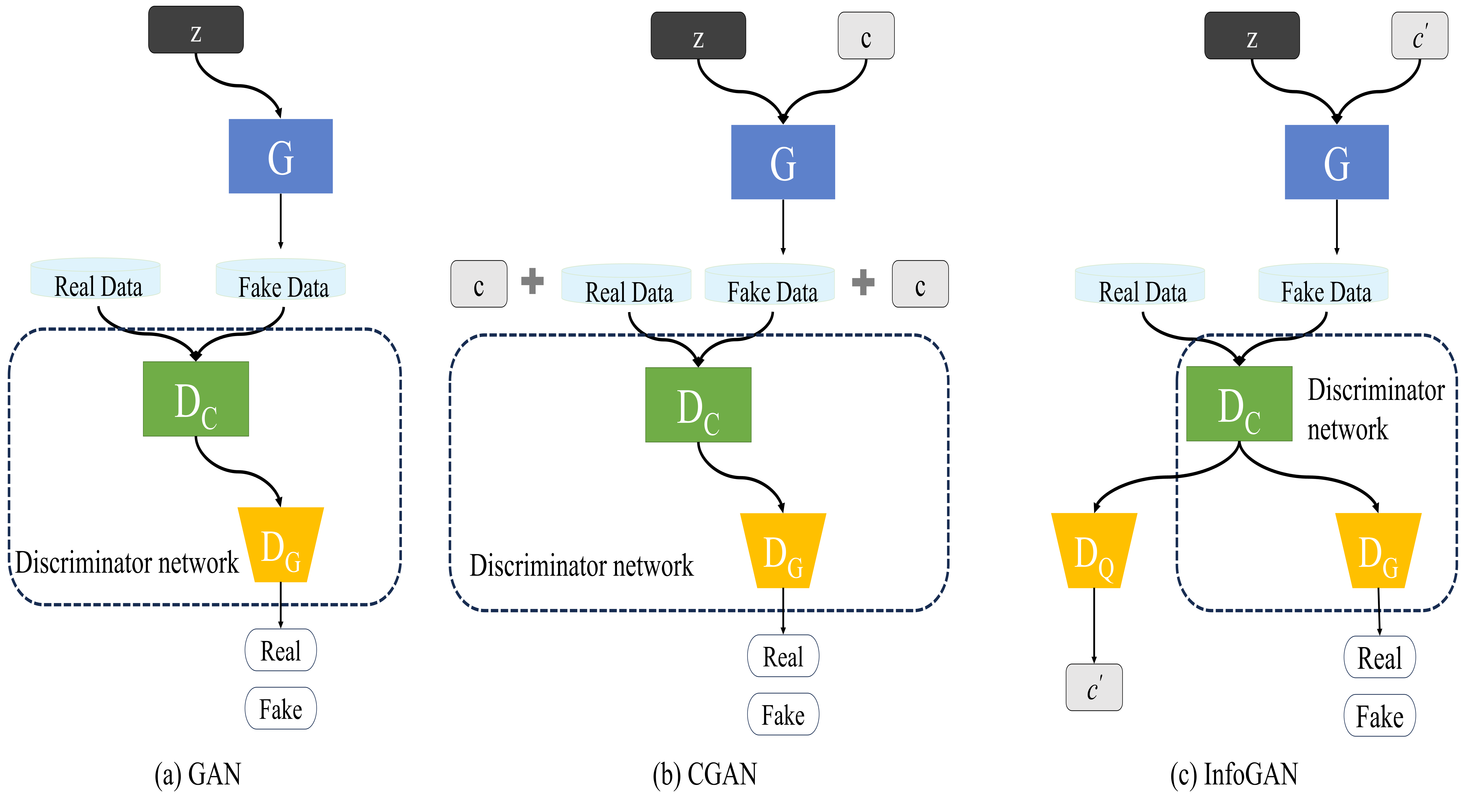}
	\begin{center}
	\caption{The construction of the models, \textbf{z} represents gaussian noise, \textbf{c} represents label, \textbf{$\mathbf{c^{'}}$} represents latent code.}
	\label{fig1}
	\end{center}
\end{figure*}

\subsection{ GAN and cGAN}
GAN consists of two parts: a generator network (G) and a discriminator network (D). The generator tries to generate diversified synthetic data G(\textbf{z}) to deceive the discriminator by learning the distribution-to-distribution mapping from Gaussian noise \textbf{z} to the actual data \textbf{x}. The discriminator tries to distinguish between G(\textbf{z}) and the real data \textbf{x} by assigning higher scores to \textbf{x} and lower scores to G(\textbf{z}). Therefore, the training process of GAN is essentially an adversarial process between generator and discriminator, which can be summarized by Fig. 1(a) and (1).
\begin{equation}
	%	\label{deqn_ex1}
	%\begin{split}
		\mathop{\min}_{G}\mathop{\max}_{D}[\mathbb{E}_{\mathbf{x}\in P_r}[logD(\mathbf{x})] 
		+ \mathbb{E}_{{\mathbf{\hat{x}}}\in P_g} [log(1-D({\mathbf{\hat{x}}}))]]
%	\end{split}
\end{equation}

\noindent where $P_r$ represents the real data distribution, $P_g$ represents the generated data. $\mathbb{E}$ is the expectation operator, and $\mathbf{\hat{x}}$=G(\textbf{z}). After the GAN model is well trained, the generator can work independently to create unlimited amount of synthetic data G(\textbf{z}). To further stabilize the GAN’s training process, Wasserstein GAN (WGAN) is proposed in \cite{arjovsky2017wasserstein} with a modified loss function:

\begin{equation}
	%	\label{deqn_ex1}
	\begin{split}
		\mathop{\min}_{G}\mathop{\max}_{D \in \omega }[\mathbb{E}_{\mathbf{x}\in P_r}[D(\mathbf{x})] 
		- \mathbb{E}_{{\mathbf{\hat{x}}}\in P_g}[D({\mathbf{\hat{x}}})]]
	\end{split}
\end{equation}

\noindent where $\omega$ is the set of 1-Lipschitz function. A gradient penalty method \cite{gulrajani2017improved} to further improve the performance of WGAN. Then the final version of the loss function becomes: 

\begin{equation}
	%	\label{deqn_ex1}
	\begin{split}
		L=\mathop{\min}_{G}\mathop{\max}_{D \in \omega}[\mathbb{E}_{\mathbf{\hat{x}}\in P_g}
		&[D(\mathbf{\hat{x}})]-\mathbb{E}_{\mathbf{x}\in P_r}[D(\mathbf{x})]]
		\\&+\lambda \mathbb{E}_{\mathbf{\widetilde{x}}\in P_{\mathbf{\widetilde{x}}}}[(\left\| \nabla_{\mathbf{\widetilde{x}}}D(\mathbf{\widetilde{x}})\right\|_2 - 1)^{2}]
	\end{split}
\end{equation}

\noindent where $\nabla$ is the gradient operator, $P_{\mathbf{\widetilde{x}}}$ is the distribution of data sampled uniformly along straight lines between pairs of points sampled from $P_r$ and $P_g$, respectively.

Note that the generator of GAN establishes the distribution-level mapping function from \textbf{z} to \textbf{x}, while ignoring the pixel-level correspondence between \textbf{z} and \textbf{x}. This makes it difficult to control and interpret the generation process. To address this issue, cGAN is proposed to generate data that satisfy given conditions. Both the generator and the discriminator in cGAN have the condition vector c (i.e., labels of the data) as an additional input to help the generator learn the conditional distribution from the actual dataset, as shown in Fig. 1(b). The loss function of cGAN is

\begin{equation}
	%	\label{deqn_ex1}
	\begin{split}
		\mathop{\min}_{G}\mathop{\max}_{D}V(D,G)=
		&\mathbb{E}_{\mathbf{x} \sim p_{data}(\mathbf{x})}[logD(\mathbf{x}|\mathbf{c})] \\
		&+ \mathbb{E}_{\mathbf{z} \sim p_{z}(\mathbf{z})}[log(1-D(G(\mathbf{z}|\mathbf{c})))]
	\end{split}
\end{equation}

As mentioned in Section I, the training of cGAN requires labels of the actual dataset, which is demanding and costly. Next we will introduce infoGAN framework that can extract interpretable features from unlabeled dataset to achieve convenient and controllable synthetic data generation. 

\subsection{InfoGAN}

In the infoGAN model, a latent code vector $\mathbf{c^{'}}$ is introduced as an additional input to the generator to represent the interpretable features to be extracted. 

If a stable correlation pattern between $\mathbf{c^{'}}$ and the generated dataset G(\textbf{z},$\mathbf{c^{'}}$) can be established, then $\mathbf{c^{'}}$ is considered to be features of G(\textbf{z},$\mathbf{c^{'}}$) with certain physical meanings. To achieve this, a mutual information term I($\mathbf{c^{'}}$,G(\textbf{z},$\mathbf{c^{'}}$)) is introduced to the GAN loss function to maximize their correlation:

\begin{equation}
	%	\label{deqn_ex1}
	\begin{split}
		\mathop{\min}_{G}\mathop{\max}_{D}V_{\mathrm{I}}(D,G)=
		V(D,G)-\lambda I(\mathbf{c^{'}},G(\mathbf{z},\mathbf{c^{'}}))
	\end{split}
\end{equation}

\noindent where $\lambda$ is the weighting factor. Note that in practice, I($\mathbf{c^{'}}$,G(\textbf{z},$\mathbf{c^{'}}$)) is hard to compute directly as it requires access to the posterior probability $P (\mathbf{c^{'}}|\mathbf{x})$ that is difficult to estimate. As an approximate solution, a classifier network Q is further introduced to approximate $P (\mathbf{c^{'}}|\mathbf{x})$, as shown in Fig. 1(c). We can define a variational lower bound, $L_{I}(G,Q)$, of the mutual information, I($\mathbf{c^{'}}$,G(\textbf{z},$\mathbf{c^{'}}$)) \cite{barber2004algorithm}. If we set the common part as  \(D_{C}\), then we have  \(D=D_{C}+D_{G}\),  \(Q=D_{C}+D_{Q}\). D and Q share the same network structure and parameters except the last layer. The generator and the classifier essentially formulate an encoder-decoder structure for the latent code. The revised loss function becomes:
 
\begin{align}
 	L_I(G, Q) &\leq I(\mathbf{c^{'}},G(\mathbf{z},\mathbf{c^{'}})) \\
 	\mathop{\min}_{G,Q}\mathop{\max}_{D}V_{\mathrm{I}}(D,G,Q) &=
 	V(D,G)-\lambda L_{I}(G,Q)
\end{align}

After the infoGAN model is trained, we can correlate the extracted latent vector  $\mathbf{c^{'}}$ with physical characteristics of the unlabeled dataset to interpret the data, as well as controlling the generation results by setting up the digits in  $\mathbf{c^{'}}$. The diversity of the generation results can be still achieved by varying Gaussian noise  $\mathbf{z}$. Meanwhile, we also apply Wasserstein distance and gradient penalty techniques to alleviate the mode collapse issue of GAN models, as mentioned in Section II. A.

\begin{figure}[!t]
	\centering
	\includegraphics[width=3in]{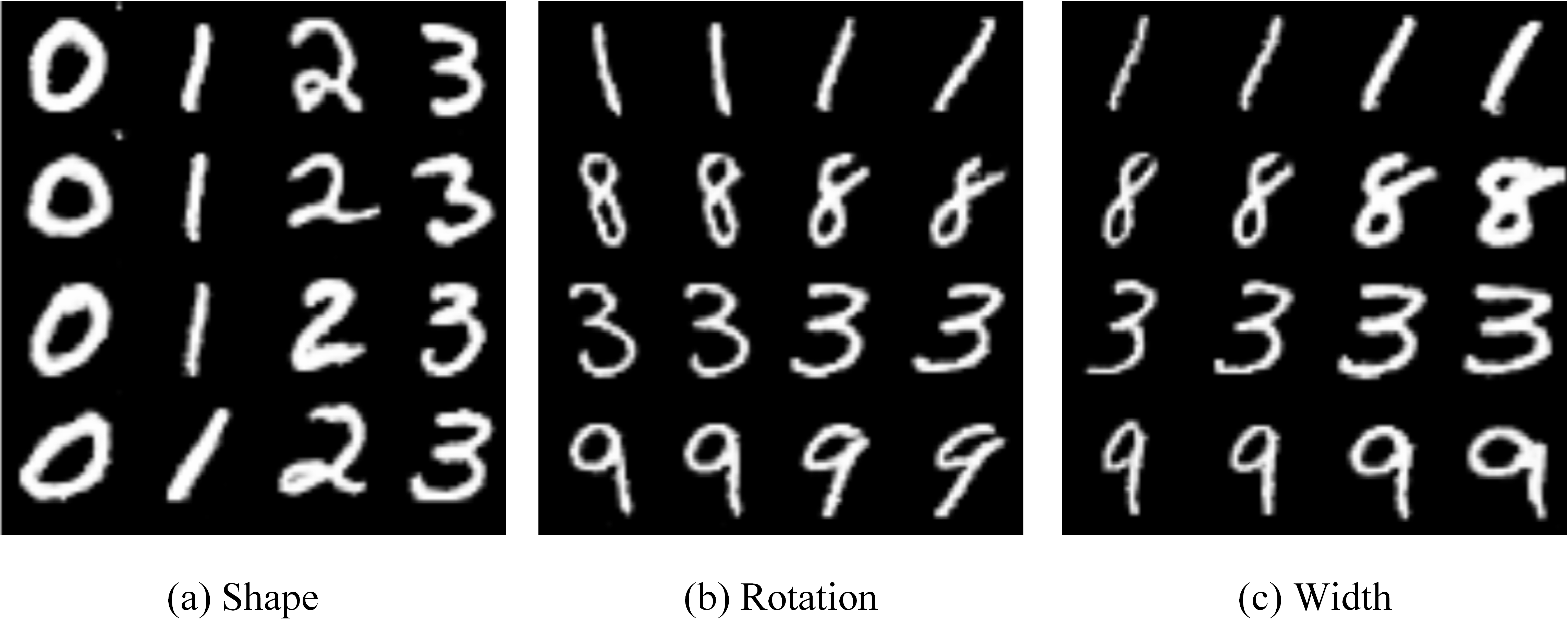}
	\caption{Manipulating latent codes on MNIST: Each latent code changes from left to right while the other latent codes are fixed.}
	\label{fig1.5}
\end{figure}

In Fig. \ref{fig1.5}, we show the results of applying the infoGAN model to the MNIST dataset. The discrete code $\mathbf{c_1}$ captures significant changes in the shape of the digits. By changing the discrete code $\mathbf{c_1}$, the type of generated digit can be altered. The continuous codes capture continuous variations. $\mathbf{c_2}$ models the rotation of the digits, and $\mathbf{c_3}$ controls the width. InfoGAN can extract interpretable features from MNIST, making the generation results controllable.

\subsection{Network Structure}

\begin{figure*}[!t]
	\centering
	\includegraphics[width=6in]{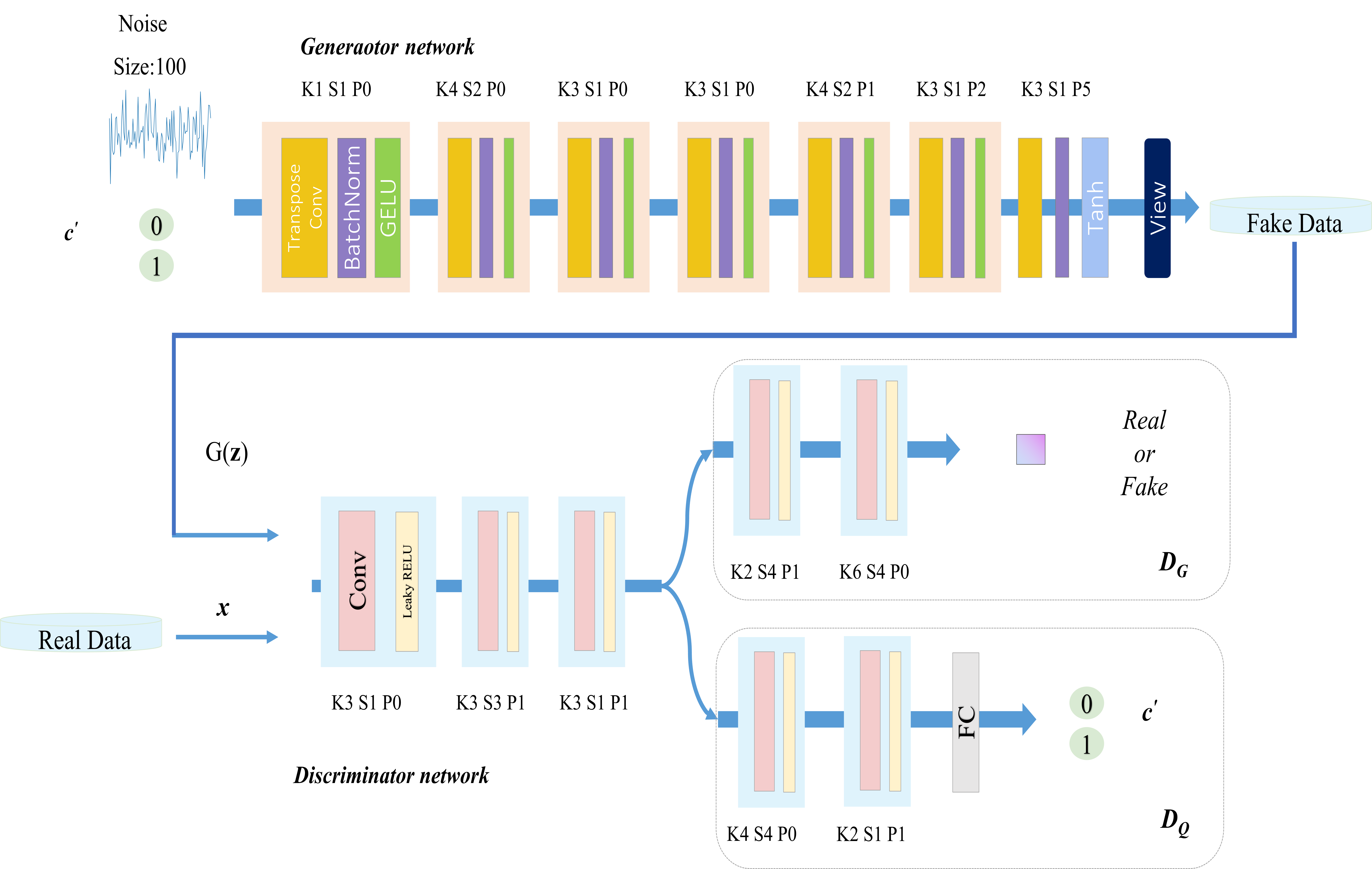}
	\caption{Realization of infoGAN with corresponding kernel size (K), stride (S), padding (P) for each convolutional layer.}
	\label{fig2}
\end{figure*}

The realization of infoGAN framework in this paper is shown in Fig. \ref{fig2}. The input is a 100-dimension Gaussian noise concatenated with a binary latent vector representing discrete features of the data. The generator network is a deep Convolutional Neural Network (CNN). Transpose convolution layers with decreasing number of kernels are implemented to convert the input vector to the generated synthetic data. Batch normalization layers \cite{ioffe2015batch} and Gaussian Error Linear Units (GELU) activation functions \cite{hendrycks2016gaussian} are attached after each convolution layer to stabilize the training process. The discriminator network is built with a set of convolution layers with increasing number of kernels and LeakyReLU activation functions. The classifier network shares the same structure and parameter with the discriminator network except the last layer to output the latent vector instead of the scores. Because Wasserstein distance and gradient penalty is implemented, there is neither batch normalization layer nor Sigmoid activation function in the discriminator network. 

\section{Case Study}
In this section, we evaluate the performance of the proposed infoGAN model in two aspects: 1) the ability to extract interpretable features from unlabeled dataset, 2) the realisticness of the generation results. Two datasets are selected to set up the test case. The first dataset is a load profile dataset including residential and industrial users from 2017 to 2019 with 15-minute granularity. We expect the infoGAN model to extract a binary feature named “load type” to correctly label these two types of load profiles, and then generate realistic synthetic load profiles for each type. The second dataset includes power output profiles of wind and photovoltaic power plants starting from 2018 to 2020 with 15-minute granularity. We first expect the infoGAN model to learn a binary feature named “power generation type” and then generate realistic synthetic profiles accordingly. In addition, in this dataset, we will also try to extract continuous features to characterize the shape (such as magnitude, volatility, etc.) of the power output profiles, thus achieving further controllability of the generation process. The model is built in PyTorch environment and trained on a single NVIDIA GeForce 3090 GPU.

\subsection{Case 1: Residential and Industrial Load Profiles}
We randomly select 700 residential daily load profiles and 700 industrial daily load profiles from the load profile dataset to set up the test case. We use the 1400 daily load profiles in total to train and evaluate the infoGAN model. The minimum-maximum normalization is implemented to preprocess the load profiles to constrain the data range to [0, 1]

\begin{equation}
	%	\label{deqn_ex1}
		x^{*} = \frac{x-x_{min}}{x_{max}-x_{min}}
\end{equation}

\noindent where $x^{*}$ is the normalized value, and $x_{max}$ and $x_{min}$ are the maximum and minimum values of the data samples, respectively. To extract a binary feature describing the load type, we set both the latent code $c^{'}$ and the output of the Q network as a two-digit vector in the one-hot encoding format. As a result, (0,1) will represent one type of load while (1,0) will represent the other. Note that due to the unsupervised learning nature of infoGAN, there is no fixed mapping between the latent code and the load type. For simplicity, we mark the load profiles labeled by (1,0) as the first class, while (0,1) as the second.

\subsubsection{Training process of infoGAN}
\begin{figure}[!t]
	\centering
	\includegraphics[width=3.5in]{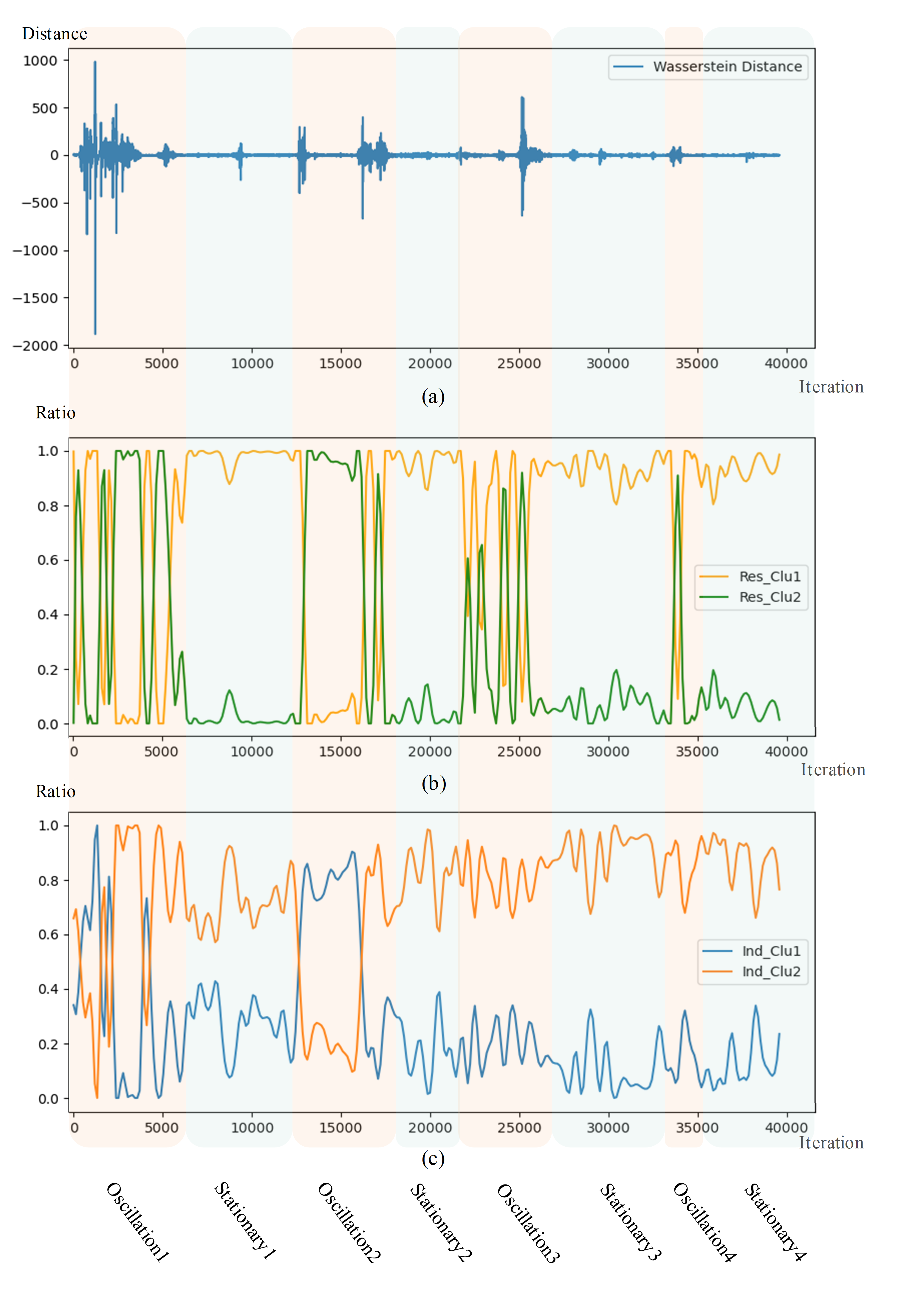}
	\caption{The training process of infoGAN, (a) Wasserstein distance (b) ratio of actualresidential load profiles assigned to class1 and class 2, (c) ratio of actual industriaload profiles assigned to class 1 and class 2.}
	\label{fig3}
\end{figure}

\ 
\newline
\indent The training process of the infoGAN model is summarized in Fig. \ref{fig3}. After every training epoch, we save the network parameters and use the classifier network Q to assign labels to each load profile sample. From Fig. \ref{fig3}, we find that the training process does not converge smoothly. Instead, there are multiple oscillation and stationary periods. During each oscillation period, the Wasserstein distance fluctuates significantly, and so do the labeling results for the load profile samples. This means the model is adjusting its parameters drastically to find a better way to label the load profiles to minimize the loss function, even if the labeling results need to be completely reversed compared to the previous epoch. After 4 oscillation and 3 stationary periods, the model tends to be stabilized and enters the final stationary period, in which most residential profiles are assigned to class 1 and most industrial profiles are assigned to class 2. In this case, once \(\mathbf{c^{'}}\) is set to (1,0), the infoGAN model will generate synthetic load profiles that has residential characteristics. 

\subsubsection{Evaluation of feature extraction results}
\begin{figure}[!t]
	\centering
	\includegraphics[width=2.8in]{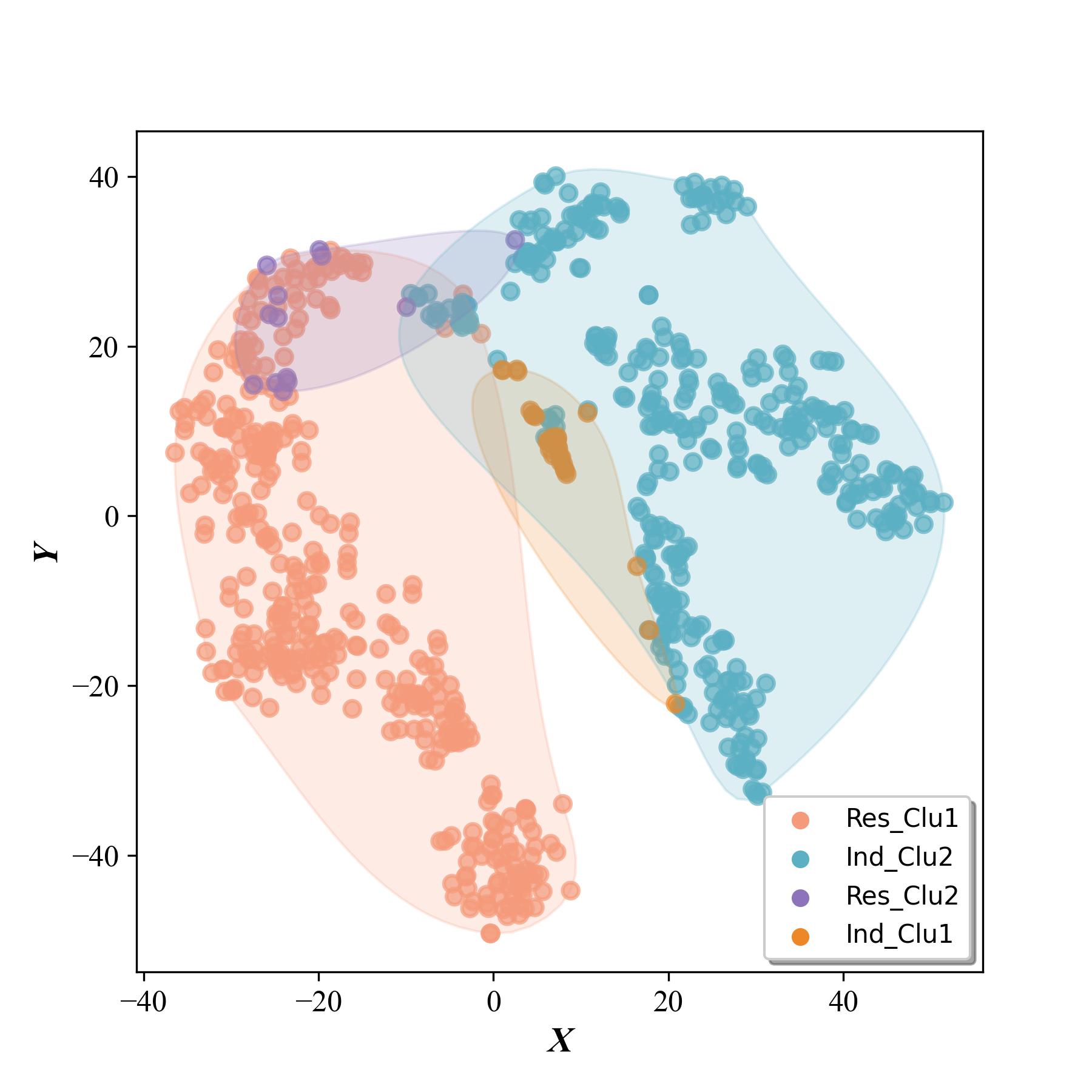}
	\caption{2-D t-SNE visualization of infoGAN results.}
	\label{fig4.7}
\end{figure}

\begin{figure}[!t]
	\centering
	\includegraphics[width=2.8in]{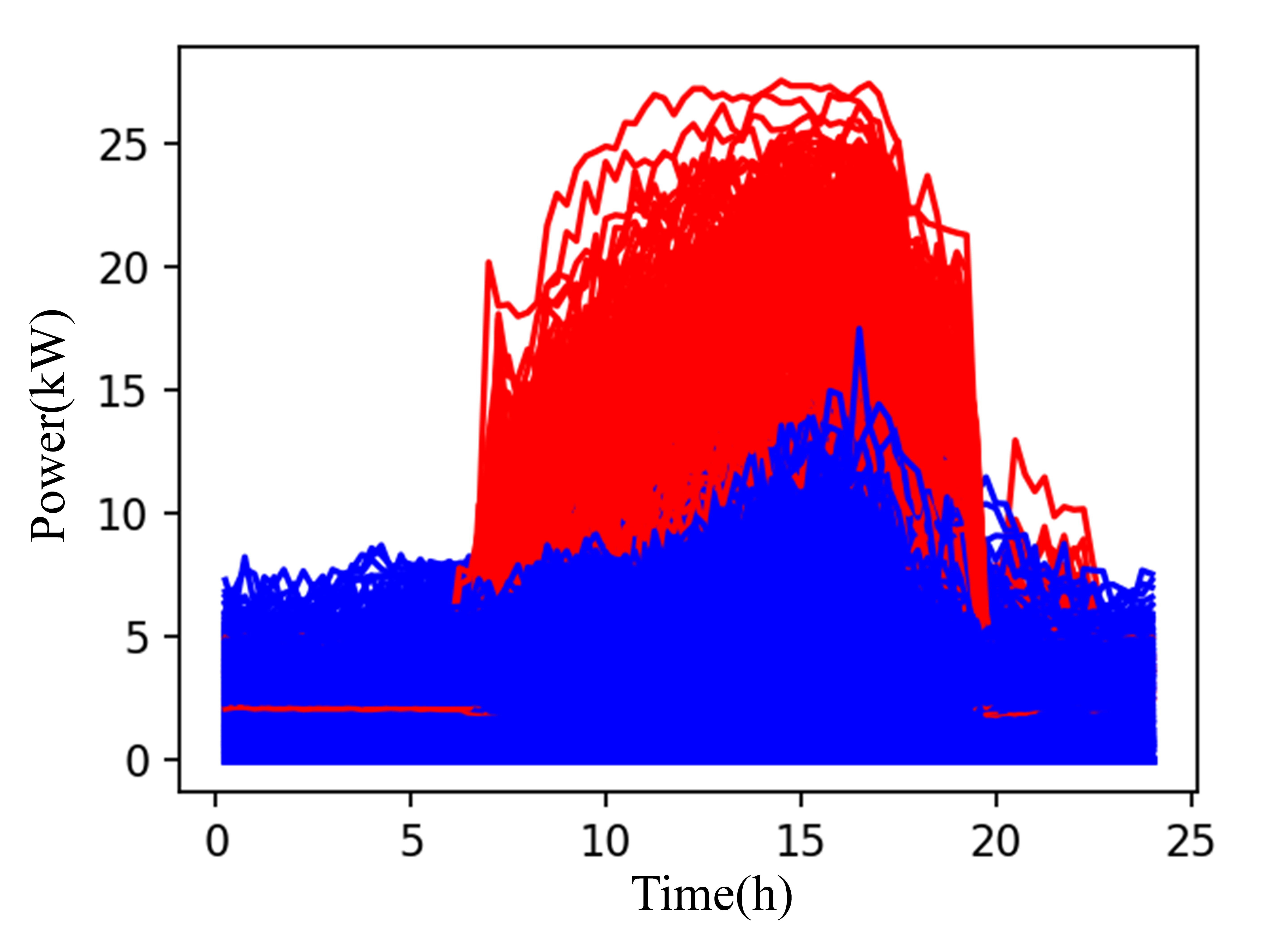}
	\caption{K-means results. Different colors represent different categories, which are basically classified according to the load level.}
	\label{fig4.5}
\end{figure}

\ 
\newline
\indent The most distinct feature of infoGAN is that it can extract interpretable features from unlabeled dataset. In our case, we expect the infoGAN model to extract a binary feature “load type” to distinguish between the residential load profiles and the industrial load profiles. If successful, the load profiles labeled (1,0) should be all residential (or industrial), while the load profiles labeled (0,1) should be all industrial (or residential). Because we have the actual load type of each load profile, we calculate the labeling accuracy after the infoGAN model is well trained, shown in Table~\ref{tab2.5}. We can see 96.31\% residential load profiles are labeled by cluster 1 with latent code (1,0), while 95.36\% industrial load profiles are labeled by cluster 2 with latent code (0,1). Fig. \ref{fig4.7} is a 2-D visualization of the load profile samples and the infoGAN labeling results by t-distributed Stochastic Neighbor Embedding (t-SNE) \cite{van2008visualizing}. We can see that the labeling error happens in the overlapping area of residential and industrial samples, the samples of which are indistinguishable in nature. 
As a comparison, we implement the k-means clustering method to classify the unlabeled load profiles into two clusters, as shown in Table~\ref{tab2.5} and Fig. \ref{fig4.5}. We can see that the overall classification accuracy of the k-means method is lower than infoGAN, especially for the industrial load. This is because k-means is a distance-based classification method that is highly sensitive to the magnitude of the load profiles. Instead, infoGAN focuses on the high-dimensional shape characteristics of the load profiles, therefore leading to better classification accuracy. More importantly, infoGAN can automatically extract a binary feature to label each load profile without human intervention. Such an auto-labeling capability is critical in processing the large amount of unlabeled data in power systems, and can be further used to achieve controllable synthetic data generation.

\begin{table}
	\renewcommand{\arraystretch}{2}
	\setlength{\tabcolsep}{10pt}
	\begin{center}
		\caption{Classification Percentage}
		\label{tab2.5}
		\begin{tabular}{|c| c | c | c |}
			\hline
			&  & Res & Ind  \\
			\hline
			\multirow{2}{*}{ InfoGAN } & Clu1 & 4.64\%  & \textbf{95.36\%} \\
			\cline{2-4}
			\multirow{2}{*}{ }    & Clu2  & \textbf{96.31\%} & 3.69\% \\
			\hline
			\multirow{2}{*}{ K-means }	& Clu1 & \textbf{100\%} & 0\% \\ 
			\cline{2-4} 
			\multirow{2}{*}{ } 	 & Clu2 & 80.75\%  & \textbf{19.25\%} \\
			\hline 
			
		\end{tabular}
	\end{center}
\end{table}

\subsubsection{Evaluation of synthetic load profile generation results}
\ 
\newline
\indent In this section, we evaluate the realisticness of the synthetic load profiles generated by infoGAN. 320 synthetic load profiles are first generated by the trained infoGAN model for both industrial and residential loads. Then we evaluate the realisticness of the generation results from two aspects: 1) we define 5 indexes to quantify and compare the shape characteristics between the real and the generated load profiles. 2) we calculate the distribution-to-distribution similarity between the real and the generated load profiles. As a benchmark, we also generate load profiles by the cGAN model that shares a similar hyperparameter configuration with infoGAN. 
The 5 indexes are defined as follows, including near-peak load, near-base load, high-load duration, rising duration and rising frequency \cite{wang2020generating,price2010methods}. 
  \begin{itemize}
      \item \textit{Near-peak load} \(\mathit{{P}_{peak}}\) is defined as the 97.5\% percentile of a load profile. We exclude the maximum value of the load profile to avoid the influence of abnormal load spikes.
      \item \textit{Near-based load} \(\mathit{{P}_{base}}\) is defined as the 2.5\% percentile of a load profile. Similarly, we exclude the minimum value to avoid the influence of zero values caused by factors such as communication failure, equipment fault, etc.
      \item \textit{High-load duration} is defined as the maximum duration of a load profile segment that is continuously larger than \(\frac{1}{2}(\mathit{{P}_{peak}} + \mathit{{P}_{base}})\), which is defined as high load.
      \item \textit{Rising duration} measures the time it takes for the load to rise from the near-base load to the high load. This index reflects the dynamic characteristics of the load profile.
      \item \textit{Rising frequency} measures how many times the load rises from the near-base load to the high load within a load profile. This index reflects the volatility of the load profile. 
  \end{itemize}

We denote the generation results of infoGAN with latent code (1,0) as GEN1, and (0,1) as GEN2. Table~\ref{tab1} shows the index calculation results for both the real and generated load profiles. We can see that GEN1 has similar characteristics with industrial loads, while GEN2 is similar with residential loads. For example, the near-peak load, near-base load and high-load duration of GEN1 is significantly larger than GEN2, showing typical characteristics of industrial loads. The rising frequency of GEN2 is higher than GEN1, indicating the volatility of residential loads. Based on Table ~\ref{tab1}, we can conclude that the generation results are statistically realistic compared with actual residential and industrial loads.

\begin{table*}
	\renewcommand{\arraystretch}{2}
	\setlength{\tabcolsep}{10pt}
	\begin{center}
		\caption{Curve Shape Indexes of Residential and Industrial}
		\label{tab1}
		\begin{tabular}{|c| c | c | c | c | c | c |c|c|}
			\hline
			\multirow{2}{*}{ Class } & \multicolumn{2}{c|}{Near-peak load} & \multicolumn{2}{c|}{Near-base load} &\multicolumn{2}{c|}{High-load duration} &
			Rising duration & Rising frequency\\
			\cline{2-9}
			\multirow{2}{*}{ } & mean & std & mean & std &  mean & std & mean & mean\\
			\hline
			Ind & 9.203 & 5.783 & 3.053 & 0.708 & 2.322 & 3.286 & 1.381 & 1.459\\
			\hline
			GEN1 & 9.542 & 6.762 & 2.058 & 1.317 & 2.796  & 3.902 & 1.468 & 1.845\\ 
			\hline
			Res & 1.156 & 0.690 & 0.434 & 0.361 & 0.627  & 1.264 & 1.789 & 3.968\\
			\hline 
			GEN2 & 1.913 & 0.911 & 0.673 & 0.233 & 1.122  & 1.562 & 1.830 & 2.462\\
			\hline 
		\end{tabular}
	\end{center}
\end{table*}

\begin{figure}[t]
	\centering
	\includegraphics[width=3.2in]{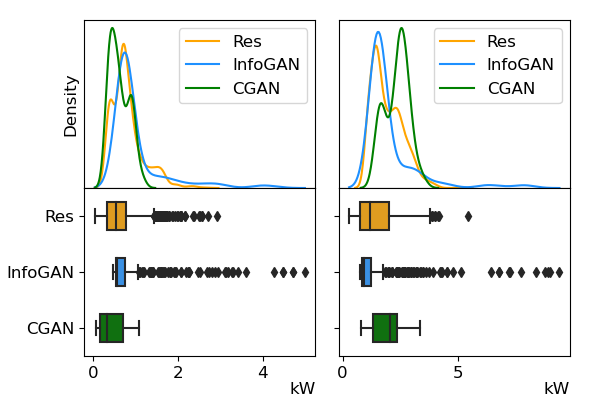}
		\begin{center}
		\footnotesize ~~~(a)Mean\qquad\qquad\qquad\quad~~~~(b)Peak
	\end{center}  
	\caption{(a) Residential mean power distribution curves and boxplots, and (b) Residential peak load value distributions and boxplots. }
	\label{fig8}
\end{figure}

\begin{figure}[t]
	\centering
	\includegraphics[width=3.2in]{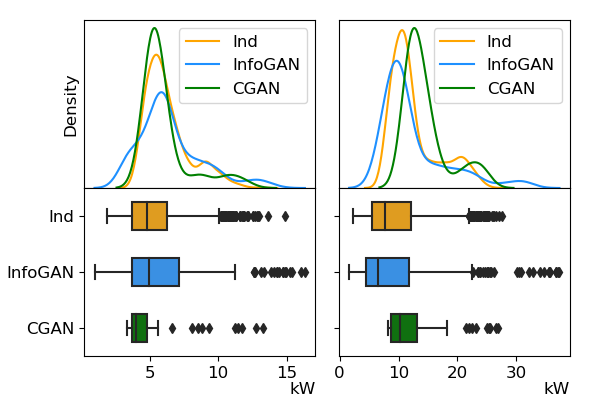}
		\begin{center}
		\footnotesize ~~~(a)Mean\qquad\qquad\qquad\quad~~~~(b)Peak
	\end{center}  
	\caption{(a) Industrial mean power distribution curves and boxplots, and (b) Industrial peak load value distributions and boxplots. }
	\label{fig9}
\end{figure}

To further evaluate the model performance, we generate synthetic residential and industrial load profiles based on cGAN. The distributions of mean and peak values of infoGAN’s results, cGAN’s results and the actual load profiles are compared in Figs. \ref{fig8}-\ref{fig9}. We can see the infoGAN and cGAN have comparable performance. InfoGAN is better at capturing the long-tail effect of the actual load distribution, while cGAN has better accuracy in approximating the main body of the actual load distribution.

We implement Kullback-Leibler divergence (KL divergence) \cite{kullback1997information} to quantify the distribution-to-distribution similarity between the actual load profiles and the generation results. Assume \(\mathbb{P}\) represents the target distribution, \(\mathbb{Q}\) represents the reference distribution. The KL divergence between \(\mathbb{P}\) and \(\mathbb{Q}\) can be calculated by

\begin{equation}
	%	\label{deqn_ex1}
	D_{KL}(\mathbb{P} \parallel \mathbb{Q}) = \sum\limits \mathbb{P} log\frac{\mathbb{P}}{\mathbb{Q}}
\end{equation}

In this paper, we set \(\mathbb{Q}\) as the distribution of actual load profiles, and \(\mathbb{P}\) as the distribution of generated load profiles. Table~\ref{tab3} summarizes the calculation results of KL divergence. Note that we calculate both \(D_{KL}(\mathbb{P} \parallel \mathbb{Q}) \) and \(D_{KL}(\mathbb{Q} \parallel \mathbb{P})\) due to the asymmetry. We can see that infoGAN has better performance (i.e., closer to the actual distribution) when generating residential load profiles, and has comparable performance when generating industrial load profiles.

\begin{table}[t]
	\renewcommand{\arraystretch}{2}
	\setlength{\tabcolsep}{10pt}
	\begin{center}
		\caption{Kullback-Leibler Divergence Analysis by Generation.}
		\label{tab3}
		\begin{tabular}{|c| c | c | c |}
\hline
&  & Res & Ind  \\
\hline
\multirow{2}{*}{ InfoGAN } & (\(\mathbb{P}\) $\parallel$ \(\mathbb{Q}\)) & 0.105 & 0.041\\
\cline{2-4}
\multirow{2}{*}{ }    & (\(\mathbb{Q}\) $\parallel$ \(\mathbb{P}\))  & 0.150 & 0.041\\
\hline
\multirow{2}{*}{ cGAN }	& (\(\mathbb{P}\) $\parallel$ \(\mathbb{Q}\))  & 0.123 & 0.042 \\ 
\cline{2-4} 
\multirow{2}{*}{ } 	 & (\(\mathbb{Q}\) $\parallel$ \(\mathbb{P}\)) & 0.163 & 0.038\\
\hline 
	\end{tabular}
	\end{center}
\end{table}

\subsection{Case 2: Photovoltaic and Wind Power Outputs}

Following the same logic in case 1, we randomly select 700 daily power output profiles for both photovoltaic power and wind power to formulate the test case. We first evaluate whether the infoGAN model can extract a binary feature to correctly label the two types of power output profiles. Then we evaluate the realisticness of the infoGAN generation results by shape indexes and distribution-level comparisons, and compare the model performance with cGAN. In addition, in this case we will further evaluate the ability of infoGAN in extracting continuous features to describe the shape characteristics of the photovoltaic power output profiles, which can enhance the controllability of the generation process.

\subsubsection{Binary feature extraction evaluation}
\ 
\newline
\indent
The labeling results of infoGAN for the daily photovoltaic and wind power output profiles are shown in Table~\ref{tab3.5}. We can see that 93.11\% of photovoltaic profiles are classified into cluster 2 with binary label (0,1), while 89.73\% of wind profiles into cluster 1 with binary label (1,0). This means infoGAN has successfully distinguished between the photovoltaic and wind profiles and established the mapping from the binary label to each type of the profiles. Such an implicit mapping learned by the neural network can be used to generate synthetic power output profiles of a given type. 

\begin{table}
	\renewcommand{\arraystretch}{2}
	\setlength{\tabcolsep}{10pt}
	\begin{center}
		\caption{Classification Percentage}
		\label{tab3.5}
		\begin{tabular}{| c | c | c |}	
			\hline
			   & Pho & Wind  \\
			\hline
		    Clu1 & 10.27\%  & \textbf{89.73\%} \\
		    \hline
			Clu2  & \textbf{93.11\%} & 6.89\% \\
			\hline
		\end{tabular}
	\end{center}
\end{table}

\subsubsection{Evaluation of synthetic power profile generation results}
\hspace{0.75em} Similar to the evaluation process in case 1, we first calculate and compare the shape indexes between the field measurement data and the generation results, as shown in Table~\ref{tab4}. We can see that infoGAN has captured the major shape characteristics for each power type. For example, the base load for the photovoltaic power is close to 0 since there is no power output during the night. The rising frequency is close to 1, because in most of the weather conditions the photovoltaic power output rises from base load to the high load only once in the morning.
 
\begin{table*}
	\renewcommand{\arraystretch}{2}
	\setlength{\tabcolsep}{10pt}
	\begin{center}
		\caption{Curve Shape Indexes of Photovoltaic and Wind}
		\label{tab4}
		\begin{tabular}{|c| c | c | c | c | c | c |c|c|}
			\hline
			\multirow{2}{*}{ Class } & \multicolumn{2}{c|}{Near-peak load} & \multicolumn{2}{c|}{Near-base load} &\multicolumn{2}{c|}{High-load duration} &
			Rising duration & Rising frequency\\
			\cline{2-9}
			\multirow{2}{*}{ } & mean & std & mean & std &  mean & std & mean & mean\\
			\hline
			Pho & 418.573 & 114.211& 0 & 0 & 1.609 &  2.041 & 1.934 & 1.047\\
			\hline
			GEN1 & 373.451& 143.892& 8.782& 9.812& 2.107 & 2.912& 1.801 & 1.144\\ 
			\hline
			Wind & 452.983& 250.949& 121.819& 137.391& 4.476& 4.736& 2.543 & 0.712\\
			\hline 
			GEN2 & 473.823 & 230.234 & 127.231 & 142.345 & 3.474 & 3.140 & 2.427& 0.813\\
			\hline 
		\end{tabular}
	\end{center}
\end{table*}

\begin{figure}[t]
	\centering
	\includegraphics[width=3.2in]{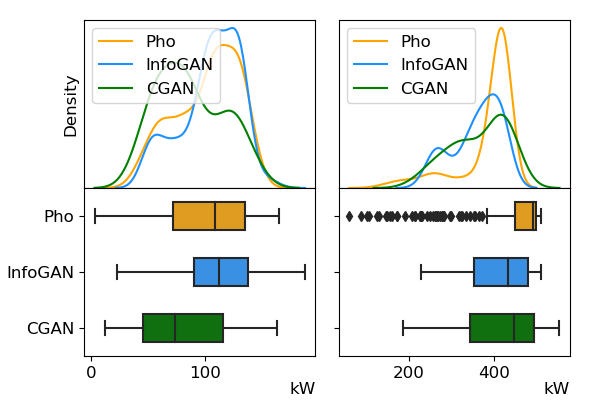}
		\begin{center}
		\footnotesize ~~~(a)Mean\qquad\qquad\qquad\quad~~~~(b)Peak
	\end{center}  
	\caption{(a) Photovoltaic mean power distribution curves and boxplots, and (b) Photovoltaic peak load value distributions and boxplots. }
	\label{fig13}
\end{figure}

\begin{figure}[t]
	\centering
	\includegraphics[width=3.2in]{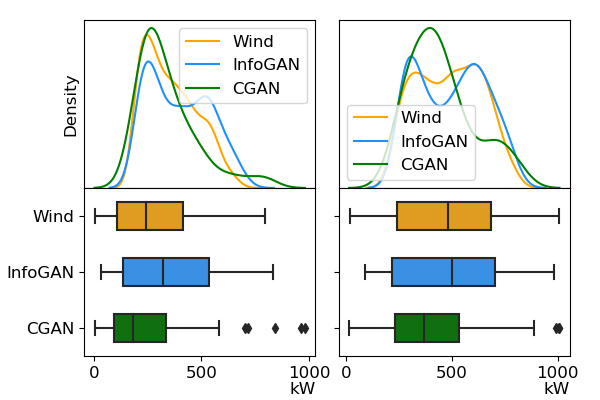}
		\begin{center}
		\footnotesize ~~~(a)Mean\qquad\qquad\qquad\quad~~~~(b)Peak
	\end{center}  
	\caption{(a) Wind mean power distribution curves and boxplots, and (b) Wind peak load value distributions and boxplots. }
	\label{fig14}
\end{figure}

\begin{table}
	\renewcommand{\arraystretch}{2}
	\setlength{\tabcolsep}{10pt}
	\begin{center}
		\caption{Kullback-Leibler Divergence Analysis by Generation.}
		\label{tab6}
		\begin{tabular}{|c| c | c | c |}
			\hline
			&  & Pho & Wind  \\
			\hline
			\multirow{2}{*}{ InfoGAN } & (\(\mathbb{P}\) $\parallel$ \(\mathbb{Q}\))) &  0.176   &  0.076\\
			\cline{2-4}
			\multirow{2}{*}{ }    & (\(\mathbb{Q}\)) $\parallel$ \(\mathbb{P}\))  &0.129  & 0.071\\
			\hline
			\multirow{2}{*}{ cGAN }	& (\(\mathbb{P}\) $\parallel$ \(\mathbb{Q}\))  & 0.161 & 0.133 \\ 
			\cline{2-4} 
			\multirow{2}{*}{ } 	 & (\(\mathbb{Q}\)) $\parallel$ \(\mathbb{P}\)) & 0.081 & 0.192\\
			\hline 
		\end{tabular}
	\end{center}
\end{table}

We also implement cGAN as a benchmark to compare the model performance, as shown in Figs. \ref{fig13}-\ref{fig14} and Table~\ref{tab6}. We can see that infoGAN achieves better approximation for the mean value distribution of photovoltaic power and the double-peak distribution characteristics of wind power. The KL divergence indicates that cGAN and infoGAN have comparable performance in general, where infoGAN performs better in wind power and cGAN performs better in photovoltaic.

\subsection{Continuous Feature Extraction Evaluation}
\begin{figure*}[!t]
	\centering
	\includegraphics[width=7in]{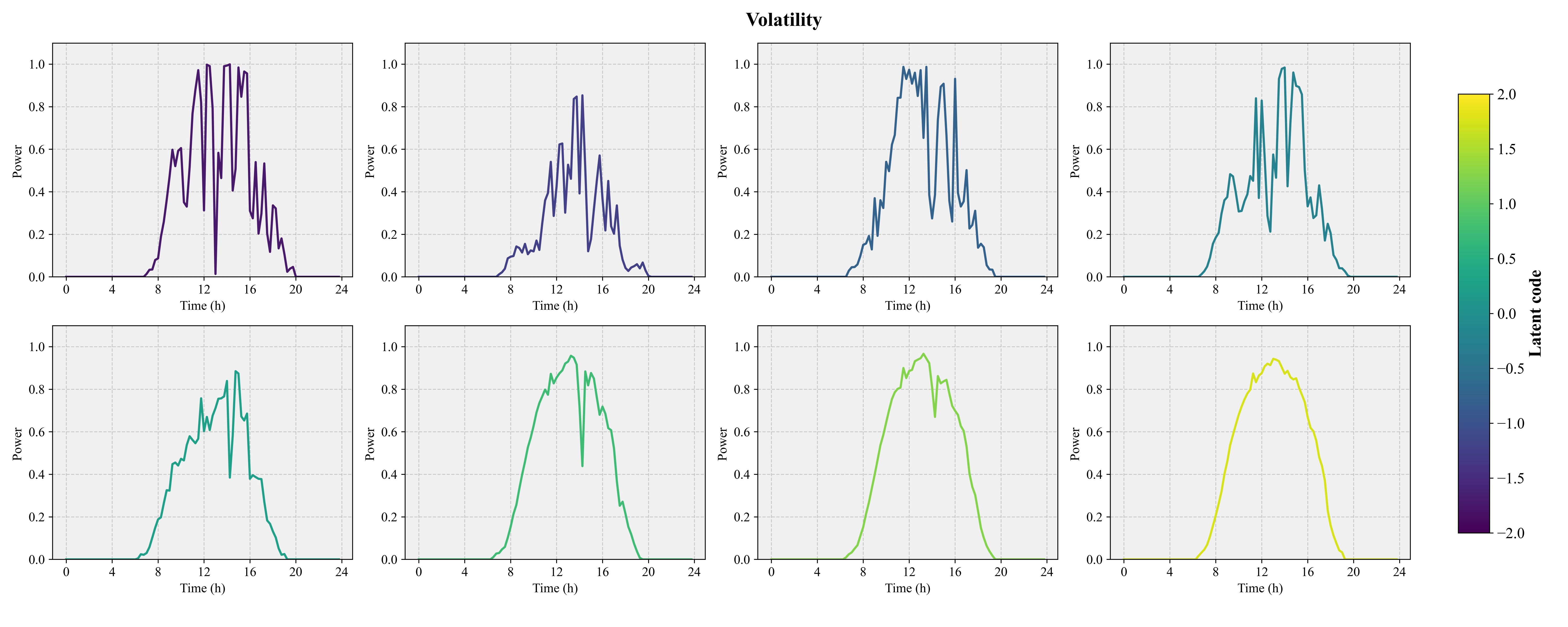}
	\caption{Schematic diagram of PV output volatility varying with continuous latent code. The sampling results of the continuous latent code generation. The continuous latent code increases sequentially from left to right and from top to bottom.}
	\label{fig:Volatility}
\end{figure*}

\begin{figure*}[!t]
	\centering
	\includegraphics[width=7in]{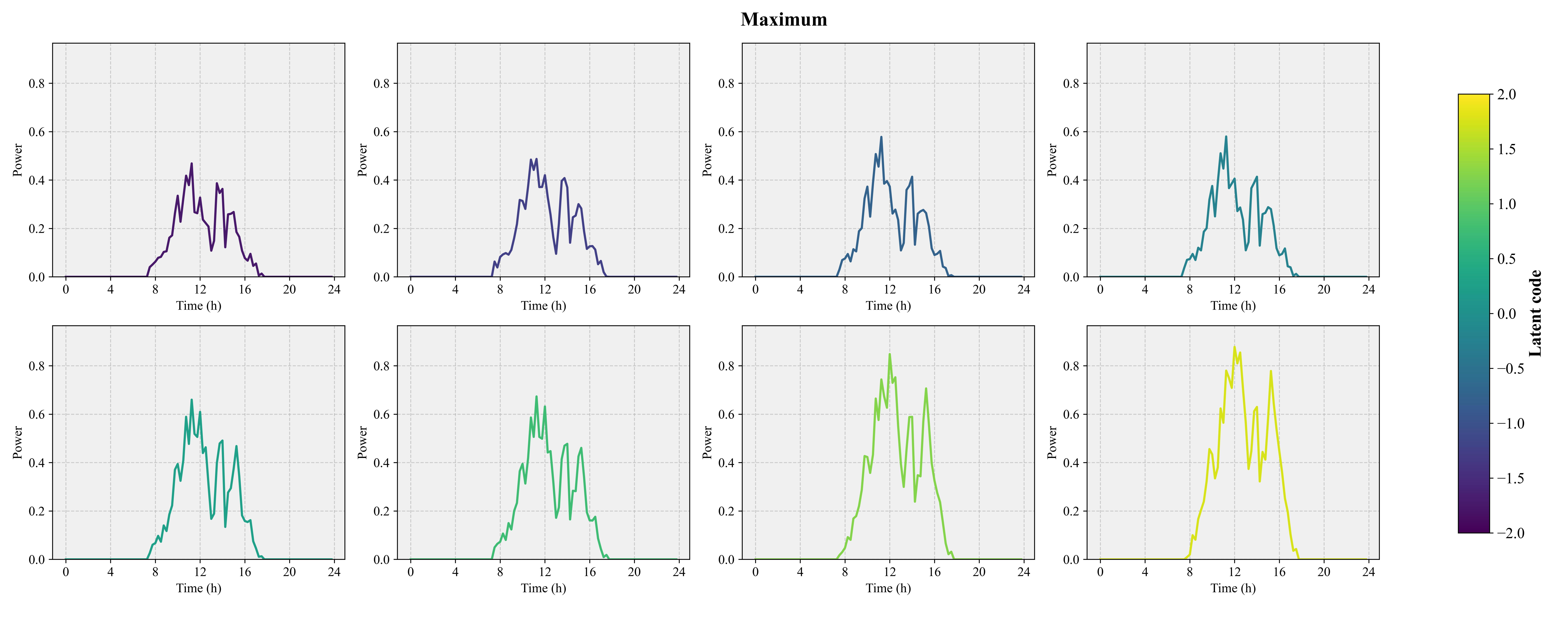}
	\caption{Schematic diagram of PV output maximum varying with continuous latent code. The sampling results of the continuous latent code generation. The continuous latent code increases sequentially from left to right and from top to bottom.}
	\label{fig:Maximum}
\end{figure*}

 In addition to case 1, in this case we further evaluate whether infoGAN can extract continuous features (such as magnitude and volatility of the profile) to better characterize the power output profiles, based on which we can achieve finer control of the follow-up synthetic data generation process. We select the photovoltaic power output profiles to formulate the test data, and concatenate a one-dimension continuous latent code to the original binary latent code to represent the continuous feature to be extracted. The continuous latent code is initialized by a uniform distribution within [-2, 2]. After the infoGAN is trained, we will generate synthetic photovoltaic power output profiles constrained by different set up of the continuous latent code to demonstrate the effectiveness. The experiment is conducted twice on two different subsets of the photovoltaic power data set, one includes all weather conditions and the other only includes cloudy days. The generation results are shown in Fig. \ref{fig:Volatility} and Fig. \ref{fig:Maximum}. 

\begin{table*}[!t]
	\renewcommand{\arraystretch}{2}
	\setlength{\tabcolsep}{10pt}
	\begin{center}
		\caption{Results of Volatility Indicator Calculation}
		\begin{tabular}{|c|c|c|c|c|c|c|c|c|}
			\hline
			Latent code & -2.000 & -1.428 & -0.857 & -0.286 & 0.286 & 0.857 & 1.428 & 2.000 \\
			\hline
			SAFOD & 9.323 & 5.892 & 6.805 & 6.349 & 3.303 & 3.154 & 2.416 & 1.967 \\
			\hline
			CV & 0.720 & 0.809 & 0.719 & 0.714 & 0.629 & 0.613 & 0.602 & 0.568  \\
			\hline
			RSAFODM & 9.334 & 6.906 &  6.895 & 6.453 & 3.733 & 3.295 & 2.500 & 2.086  \\
			\hline
		\end{tabular}
		\label{table:Volatility}
	\end{center}
\end{table*}

From Figs. \ref{fig:Volatility}-\ref{fig:Maximum}, we can see that the continuous latent code is highly correlated with a certain shape characteristic of the photovoltaic profiles. More specifically, in Fig. \ref{fig:Volatility} when the continuous latent code increases from -2 to 2, the generated profiles share similar magnitude but become more and more smooth. As a result, we can conclude that the continuous latent code represents the volatility of the photovoltaic power profiles, which is granted by interpretable physical meanings. To better quantify the volatility of the profiles, we define 3 indexes including Sum of Absolute First-order Difference (SAFOD), Coefficient of Variation (CV) and Ratio of the Sum of Absolute First-order Difference to the Maximum (RSAFODM):

\begin{equation}
	%	\label{deqn_ex1}
	\text{SAFOD} = \sum_{i=1}^{N-1} |x_{i+1} - x_i|
\end{equation}

\begin{equation}
	%	\label{deqn_ex1}
	\text{CV} = \frac{\sigma}{\mu}
\end{equation}

\begin{equation}
	%	\label{deqn_ex1}
	\text{RSAFODM} = \frac{\sum_{i=1}^{N-1} |x_{i+1} - x_i|}{\max(x)}
\end{equation}

\noindent where \(N\) is the sequence length, \( x_i \) is the \(i^{th}\) element in the sequence, \( \sigma \) and \( \mu \) are the standard deviation and mean value of the sequence, respectively. The index calculation results are shown in Table~\ref{table:Volatility}. We can see that all 3 indexes show clear decreasing trend along with the increase of the continuous latent code, indicating that the latent code successfully describes the profile volatility.

Similar conclusions can be obtained in Fig. \ref{fig:Maximum}. We can see that under cloudy days, the magnitude of the generated profiles increases along with the continuous latent code increasing from -2 to 2, while the profile volatility remains stable. This indicates that the continuous latent code represents the magnitude of the profiles. Also, the continuous latent code will be granted with different physical meanings when the infoGAN is trained on different datasets. 

\section{Conclusion}
In this paper, we introduce infoGAN model to achieve interpretable feature extraction and controllable synthetic data generation for unlabeled electrical time series data. Compared to classical GAN-based methods such as cGAN, the most distinct characteristic of infoGAN is that 
an additional latent code is included in the model input to act as the data features to be extracted, and a classifier sharing the same parameters with the discriminator is introduced to output the latent code. Moreover, a mutual information term is included in the loss function to maximize the correlation between the input latent code and the classifier output. As a result, the generator and the discriminator become an encoder-decoder structure, forcing the latent code to become representative features of the unlabeled dataset, which usually have clear physical meanings and can be further used to control the generation results.
We have the following 3 key observations from the case study results: 1) The infoGAN training process has multiple oscillation and stationary periods, which is different from classical GAN-based methods. However, such a training process is considered reasonable because the infoGAN model is seeking for an optimal way of labeling the data samples, which may lead to drastic changes of both the labeling results and the network losses. 2) InfoGAN can extract both discrete features (load type, renewable power generation type) and continuous features (profile magnitude and volatility) of the electrical time series. 3) The generation results of infoGAN has comparable realisticness with classical GAN-based methods such as cGAN.
Future work may focus on increasing the dimension of the input latent code to extract multiple interpretable features from the unlabeled dataset. 

%\begin{thebibliography}{1}

\bibliography{IEEEabrv, myrefs}
%\end{thebibliography}

\end{document}